\documentclass[aps,prb,preprint,superscriptaddress,a4paper]{revtex4}
\usepackage{graphicx}
\usepackage{amsmath}
\usepackage{mathrsfs}
\begin{document}
\title{\bf Atomistic spin dynamics of the CuMn spin glass alloy}

\author{B. Skubic}
\affiliation{Department of Physics and Materials Science, Uppsala
University, Box 530, SE-751 21 Uppsala, Sweden}

\author{O. E. Peil}
\email[]{Oleg.Peil@fysik.uu.se} \affiliation{Department of Materials
Science and Engineering, KTH, SE-100 44 Stockholm, Sweden}
\affiliation{Department of Physics and Materials Science, Uppsala
University, Box 530, SE-751 21 Uppsala, Sweden}

\author{J. Hellsvik}
\affiliation{Department of Physics and Materials Science, Uppsala
University, Box 530, SE-751 21 Uppsala, Sweden}

\author{P. Nordblad}
\affiliation{Solid State Physics, Department of Engineering
Sciences, Uppsala, Box 534, SE-751-21, Sweden}

\author{L. Nordstr\"{o}m}
\affiliation{Department of Physics and Materials Science, Uppsala
University, Box 530, SE-751 21 Uppsala, Sweden}

\author{O. Eriksson}
\affiliation{Department of Physics and Materials Science, Uppsala
University, Box 530, SE-751 21 Uppsala, Sweden}

\begin{abstract}
We demonstrate the use of Langevin spin dynamics for studying
dynamical properties of an archetypical spin glass system.
Simulations are performed on CuMn (20\% Mn) where we study the
relaxation that follows a sudden quench of the system to the low
temperature phase. The system is modeled by a Heisenberg Hamiltonian
where the Heisenberg interaction parameters are calculated by means
of first-principles density functional theory. Simulations are
performed by numerically solving the Langevin equations of motion
for the atomic spins. It is shown that dynamics is governed, to a
large degree, by the damping parameter in the equations of motion
and the system size. For large damping and large system sizes we
observe the typical aging regime.
\end{abstract}

\maketitle

\section{Introduction}
Spin glasses exhibit exotic dynamical properties such as aging,
memory, and rejuvenation, which have triggered a lot of research in
the past.\cite{Binder1986, Youngbook} The interest in dynamics in
these systems is mainly motivated by the fact that practically only
off-equilibrium properties of spin glasses can be observed in
experiments. Peculiarities in the dynamical behavior are often
related to a complex nature of the phase space and the lack of
ergodicity in spin-glass systems.\cite{Palmer_1982} In particular,
relaxation towards equilibrium is not characterized by a single
timescale but rather by a broad spectrum of relaxation times leading
to a non-trivial evolution of measurable quantities, like e.g.
magnetization.

An aging experiment is an elegant way to reveal the multiscale
nature of the spin-glass dynamics. \cite{Lundgren_1983} A system is
prepared by quenching from a high temperature to a given temperature
$T$ and perturbing the system in some way, usually by switching on
an external magnetic field. A measurement of the time evolution of
the magnetization is performed after a certain time, $t_w$, has
passed since the system preparation. For temperatures $T$ below the
spin-glass transition temperature $T_g$, relaxation of the
magnetization towards the equilibrium value shows a strong
dependence on the waiting time, $t_w$. A number of phenomenological
models has been proposed to explain this behavior, among which are
the well-known droplet model \cite{Fisher1988} and a class of
hierarchial models. \cite{Ogielski_1985b, Teitel_1985,
Huberman_1985, Sibani_1989, Bouchaud_1995} However, scaling laws
resulting from these models does not allow to interpret experimental
results unambiguously, since this would require the access to
asymptotic regimes of relaxation and hence enormously large time
scales. Moreover, different, sometimes even contradicting, models
give rise to same scaling laws, discrediting their predictions.
Studying realistic models seems therefore to be a more promising way
to elucidate mechanisms underlying spin-glass dynamics.

There has been much work on the non-equilibrium behavior of
Edward-Anderson (EA) spin-glass models. \cite{Andersson_1992,
Ozeki_2001, Katzgraber2005, Blundell1992, Ogielski_1985a,
Picco_2001, Rieger1993, Berthier2002} These numerical studies have
shown that the dynamical behavior of the correlation function is
very similar to that of the magnetization (related to the
correlation function via the fluctuation-dissipation theorem in
equilibrium) in experiments. In particular, the two-stage relaxation
depending on the waiting time is reproduced in the simulations.

In this paper, we study the spin-glass dynamics of the
Cu$_{80}$Mn$_{20}$ alloy by modeling it with the random-site
Heisenberg Hamiltonian:
\begin{equation}
\mathscr{H} = -\sum_{i,j}J_{ij}c_{i}c_{j}\textbf{m}_{i}\cdot\textbf{m}_{j},
\label{eq:hamiltonian}
\end{equation}
where $\textbf{m}_{i}$ represent vector magnetic moments and $c_i$
are the occupation numbers of the magnetic atoms ($c_i$ is equal to
1 if a site is occupied by a Mn atom and 0 otherwise); exchange
parameters $J_{ij}$ are calculated accurately within the density
function theory (DFT) approach. Usually, dynamics of spin glasses is
studied by means of Monte-Carlo simulations. \cite{Berthier2002,
Blundell1992, Rieger1993, Katzgraber2005, Kisker1996,
Ogielski_1985a} Such an approach misses some details of local
relaxations of spins in their local fields. The influence of the
finite rate of these local relaxations on the dynamics in general
can be investigated by solving the Langevin dynamics equations
directly. As a matter of fact, as will be shown below, motion of
individual spins can have a rather strong impact on the aging
behavior of the system.

In the next section, we introduce definitions used throughout the
paper and describe briefly the way aging is observed in experiments.
In Section~\ref{section.methods}, governing equations for the
atomistic spin dynamics are given along with some details on the
implementation. Results of the numerical simulations of the CuMn
alloy and EA model are presented in Section~\ref{section.results}.
The focus is on the influence of damping on the spin dynamics.

\section{Definitions and Theory}

In a typical aging experiment, a system is quenched in zero field
from high temperature to a temperature below $T_g$. Then the system
is aged during a waiting time $t_w$, a small constant field $h$ is
applied and the time dependence of the magnetization $M(t)$ or
susceptibility $\chi(t)=M(t)/h$ is observed. The relaxation process
after the quench can be associated with three phases: (1) an initial
relaxation towards a local quasi-equilibrium state, (2) aging
dynamics and (3) global equilibration. The latter is achievable only
for systems of finite size $N$ within a time greater than the
ergodic time $\tau_{erg}\sim \exp{(N)}$. Although equilibration is
of little interest in an experiment, it must be taken into account
in numerical simulations when one deals with relatively small
systems.

In numerical simulations, it is convenient to work with the
autocorrelation function:
\begin{equation}
C(t_w+t,t_w)=\frac{1}{N}\sum_{i}\left[\textbf{m}_i(t_w)\cdot\textbf{m}_i(t_w+t)
\right]_{\textrm{av}},
\end{equation}
where $[\dots]_{\textrm{av}}$ stand for configuration averaging,
i.e. averaging over independent runs with randomly generated atomic
distributions $\{c_i\}$. In the quasi-equilibrium phase, when the
conditions for the fluctuation-dissipation theorem (FDT) are
fulfilled, the autocorrelation is related to the response function
\cite{Youngbook}:
\begin{equation}
R(t_w+t,t_w) = -\frac{1}{T}\frac{\partial C(t_w+t,t_w)}{\partial t_w}.
\end{equation}

Within linear response, the (thermoremanent) magnetization and
susceptibility can be found in the following way:
\begin{eqnarray}
M(t_w+t,t_w) & = & h \int_{0}^{t_w} dt' R(t_w+t,t'), \\
\chi(\omega,t_w) & = & \int_{0}^{t_w} dt' R(t_w,t') e^{i\omega(t'-t_w)},
\end{eqnarray}
where $h$ is a small applied magnetic field in a corresponding
thermoremanent magnetization (TRM) experiment. In quasi-equilibrium,
the relaxation of observables does not depend on $t_w$ and the
relation between the magnetization and the autocorrelation function
amounts to
\begin{equation}
M(t) = M_0 - \frac{h}{T}
C_{\mathrm{eq}}(t),
\label{eq:mag}
\end{equation}
where $C_{\mathrm{eq}}(t)\equiv C(t_w+t,t_w)$ in a $t_w$-independent
regime. In a non-equilibrium situation, when the FDT is violated,
different stages of the autocorrelation relaxation can be
illustrated by the relaxation rate function defined as
\begin{equation}
S(t_w+t,t_w)  =  -\frac{d}{d\ln{t}} C(t_w+t,t_w).
\label{eq:relax}
\end{equation}
The relaxation rate peaks at a time $t_a$ of the order of the age of
the system, i.e. $t_a \sim t_w$.

Spin dynamics of a spin glass is essentially non-equilibrium, and
the aging behavior in particular makes the evolution of the
autocorrelation function be strongly dependent on the waiting time,
i.e. time-translation invariance is violated. However, under certain
circumstances or during certain time intervals, a condition of
time-translation invariance, $C(t_w+t,t_w)=C(t)$, holds and the
relaxation of the system is said to proceed in the quasi-equilibrium
regime. This assertion can be considered as a definition of a
quasi-equilibrium state. In order for the true equilibrium dynamics
to take place, the fluctuation-dissipation theorem must hold.

Having been prepared at a temperature $T$, the system tends to the
equilibrium state. This state can be characterized by the space
correlation function which for spin-glasses is calculated in the
following way:
\begin{equation}
G(R)=\frac{1}{N}\sum_i \left[(\langle \mathbf{m}_i\cdot\mathbf{m}_{i+R}\rangle-\langle
\mathbf{m}_i\rangle\cdot \langle \mathbf{m}_{i+R}\rangle)^2\right]_{\textrm{av}}.
\label{eq:scorr}
\end{equation}
Above the transition temperature for sufficiently large $R$, the
spatial correlation is given by
\begin{equation}
G(R)\sim R^{d-2+\eta}u(R/\xi),
\end{equation}
where $d$ is the dimension of the system, $\eta$ a critical
exponent, $\xi$ the correlation length and $u(x)$ a scaling function
which decays to zero for $R/\xi \rightarrow \infty$. In the
macroscopic limit, the correlation length $\xi$ is finite above
$T_g$ but diverges with $\xi \sim \epsilon^{-\nu}$ as $T_g$ is
approached from above, where $\epsilon=(T-T_g)/T_g$ is the reduced
temperature and $\nu$ is a critical exponent.

\section{Description of method}
\label{section.methods}

Our simulations are performed using the ASD (Atomic Spin Dynamics)
package\cite{Skubic2008} which is based on an atomistic approach for
spin dynamics. We use a parametrization of the interatomic exchange
part of the Hamiltonian in the form of Eq.~\ref{eq:hamiltonian}. The
effect of temperature is modeled by Langevin dynamics. Connection to
an external thermal bath is modeled with a Gilbert-like damping
characterized by a damping parameter $\alpha$.

The microscopic equations of motion for the atomic moments,
$\mathbf{m}_i$, in an effective field, $\mathbf{B}_{i}$, are
expressed as follows:
\begin{equation}
\frac{d\mathbf{m}_i}{dt}=-\gamma \mathbf{m}_i \times [\mathbf{B}_{i}+\mathbf{b}_{i}(t)]-\gamma \frac{\alpha}{m} \mathbf{m}_i \times (\mathbf{m}_i \times [\mathbf{B}_{i}+\mathbf{b}_{i}(t)]).
\label{eq:sllg}
\end{equation}
In this expression $\gamma$ is the electron gyromagnetic ratio and
$\mathbf{b}_{i}(t)$ is a stochastic magnetic field with a Gaussian
distribution. The magnitude of that field is related to the damping
parameter, $\alpha$, in order for the system to eventually reach
thermal equilibrium.

The effective field, $\mathbf{B}_i$, on a site $\it i$, is
calculated from
\begin{equation}
\mathbf{B}_i=-\frac{\partial \mathscr{H}}{\partial \mathbf{m}_i},
\label{eq:heisenberg}
\end{equation}
where for $\mathscr{H}$ we use the classical Heisenberg Hamiltonian
defined by Eq.~\ref{eq:hamiltonian}.

We use Heuns method (for details see Ref.~\onlinecite{Skubic2008})
with a time step size of 0.01 femtoseconds for solving the
stochastic differential equations. Most of the calculations are
performed up to a time $t=70 \mathrm{ps}$.

\section{Langevin spin dynamics of a Heisenberg spin glass}
\label{section.results}

\subsection{Spin dynamics of CuMn}

Spin dynamics simulations are performed for the CuMn alloy with 20\%
of magnetic atoms (Mn). The system is described by the Heisenberg
Hamiltonian with spins (magnetic atoms) distributed over the fcc
lattice. The magnetic exchange (for the Heisenberg Hamiltonian)
parameters are obtained by means of the screened generalized
perturbation method (SGPM)\cite{Ruban2004} implemented within the
exact muffin-tin orbital (EMTO)\cite{Vitos2001} scheme. The coherent
potential approximation (CPA) and disordered local moments (DLM) are
used to treat the disordered CuMn alloy in the paramagnetic state
properly. With this model and for 20 \% Mn, critical slowing down
occurs close to the experimental freezing temperature, $T_g$, which
is approximately 90~K.\cite{Gibbs1985} In contrast to the EA model,
where bonds between atomic spins are random and the atomic sites
ordered, for CuMn, the magnetic exchange parameters between the Mn
atoms are site-independent and the atomic sites for these atoms are
randomly distributed in the lattice. Simulations are performed on a
$32\times 32 \times 32$ fcc system with 20 \% of the atomic sites
occupied by Mn. We simulate the relaxation process following a
quench from completely random spin orientations to 10~K. Averaging
is performed over 10 random alloy configurations with fixed
interatomic exchange parameters.

The main concern of the current work is to investigate the influence
of damping on the aging behavior and on spin relaxation in general,
motivation being that the damping parameter is the only parameter of
the model not derived from first principles calculations.

First, it is worth considering two limiting cases: $\alpha=0$ and
$\alpha=\infty$. The first case is trivial and corresponds to an
utterly deterministic evolution with the total energy conserved. We
can call this a "microcanonical dynamics" for brevity. Since there
is no coupling to a heat bath, the system will never reach
equilibrium in this type of dynamics. In case of the infinite
damping parameter, on the other hand, relaxation of a spin to
equilibrium with respect to its local magnetic field occurs
instantaneously, and spin dynamics becomes equivalent to the
dynamics of the heat-bath Monte Carlo method.\cite{Olive_1986} For
intermediate values of the damping parameter, we expect a system to
cross over from the initial off-equilibrium dynamics to the regime
of relaxation towards a (quasi-)~equilibrium state. The duration of
the crossover must be dependent on the damping parameter.

\begin{figure}
\includegraphics*[width=0.45\textwidth]{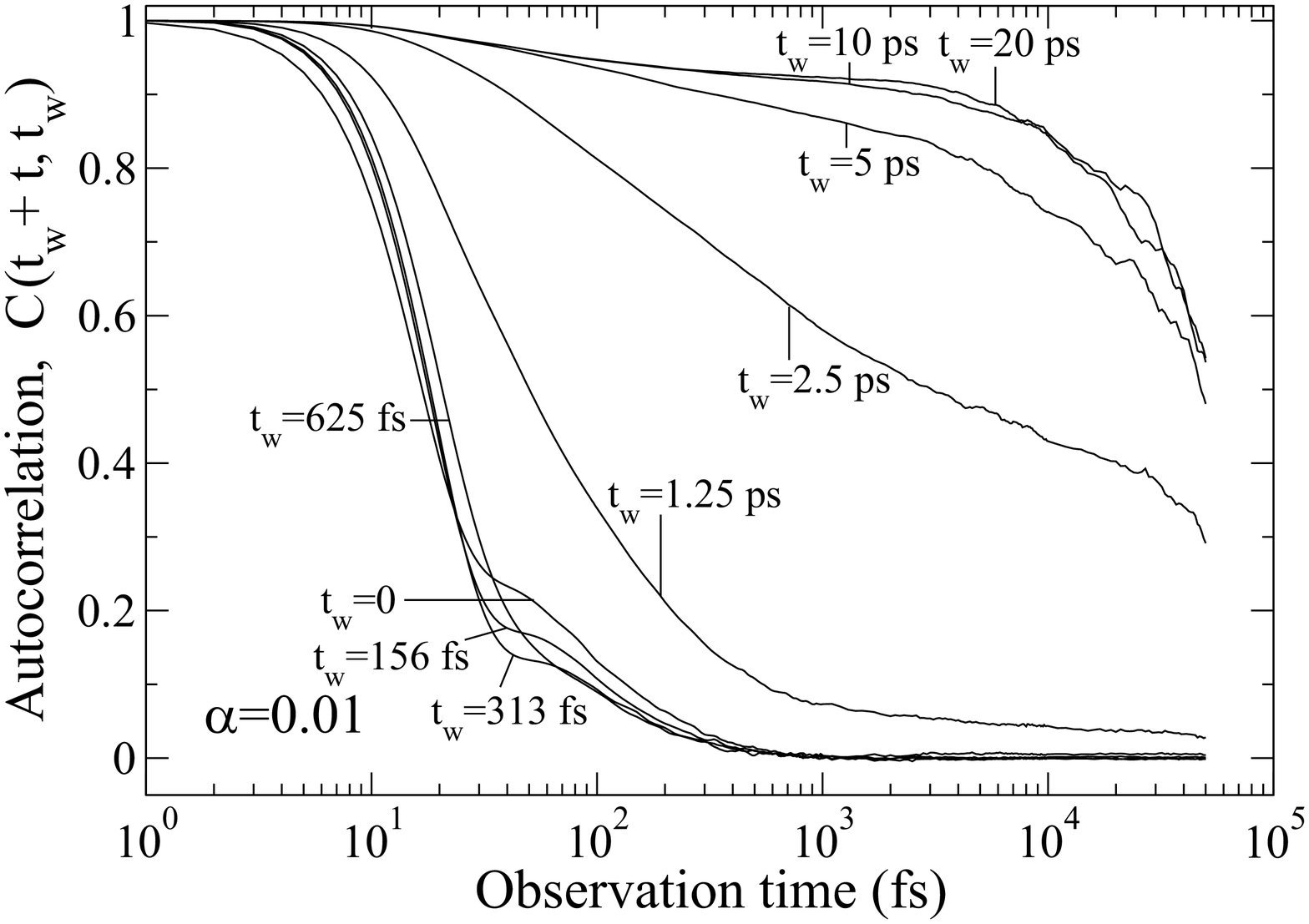}
\caption{Autocorrelation $C(t_w+t, t_w)$ calculated for CuMn after a quench
from completely random spin orientations to T=10~K with damping $\alpha=0.01$.
The autocorrelation is presented (from left to right in the figure) for
the logarithmically spaced waiting times $t_w=0$, $1.56\cdot 10^2$, $3.13\cdot 10^2$,
$6.25\cdot 10^2$, $1.25\cdot 10^3$, $2.5\cdot 10^3$, $5\cdot 10^3$, $1\cdot 10^4$, and $2\cdot 10^4$~fs.}
\label{fig:cumn1}
\end{figure}
\begin{figure}
\includegraphics*[width=0.45\textwidth]{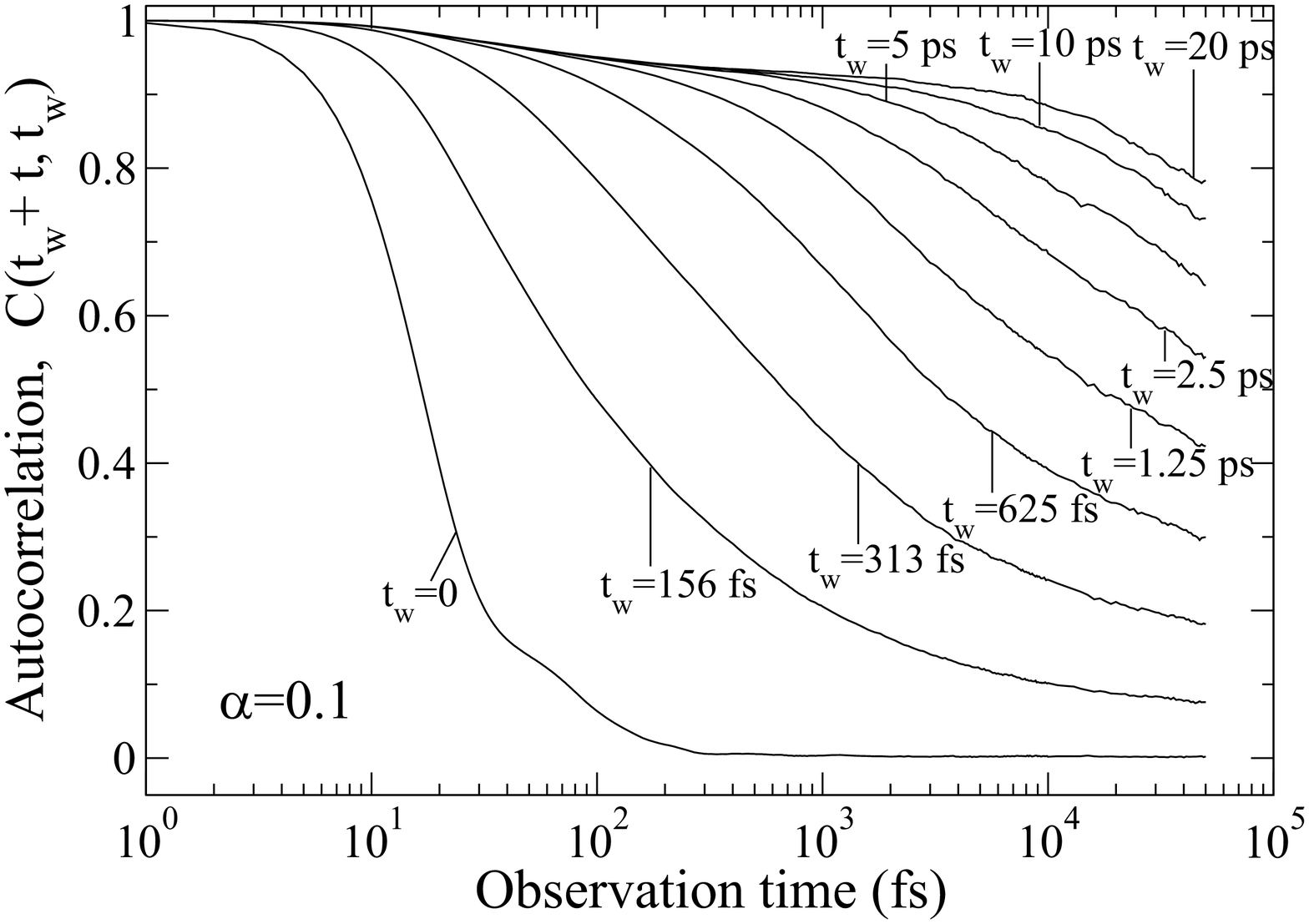}
\caption{Same as Fig.~\ref{fig:cumn1} but with damping $\alpha=0.0316$.}
\label{fig:cumn2}
\end{figure}
\begin{figure}
\includegraphics*[width=0.45\textwidth]{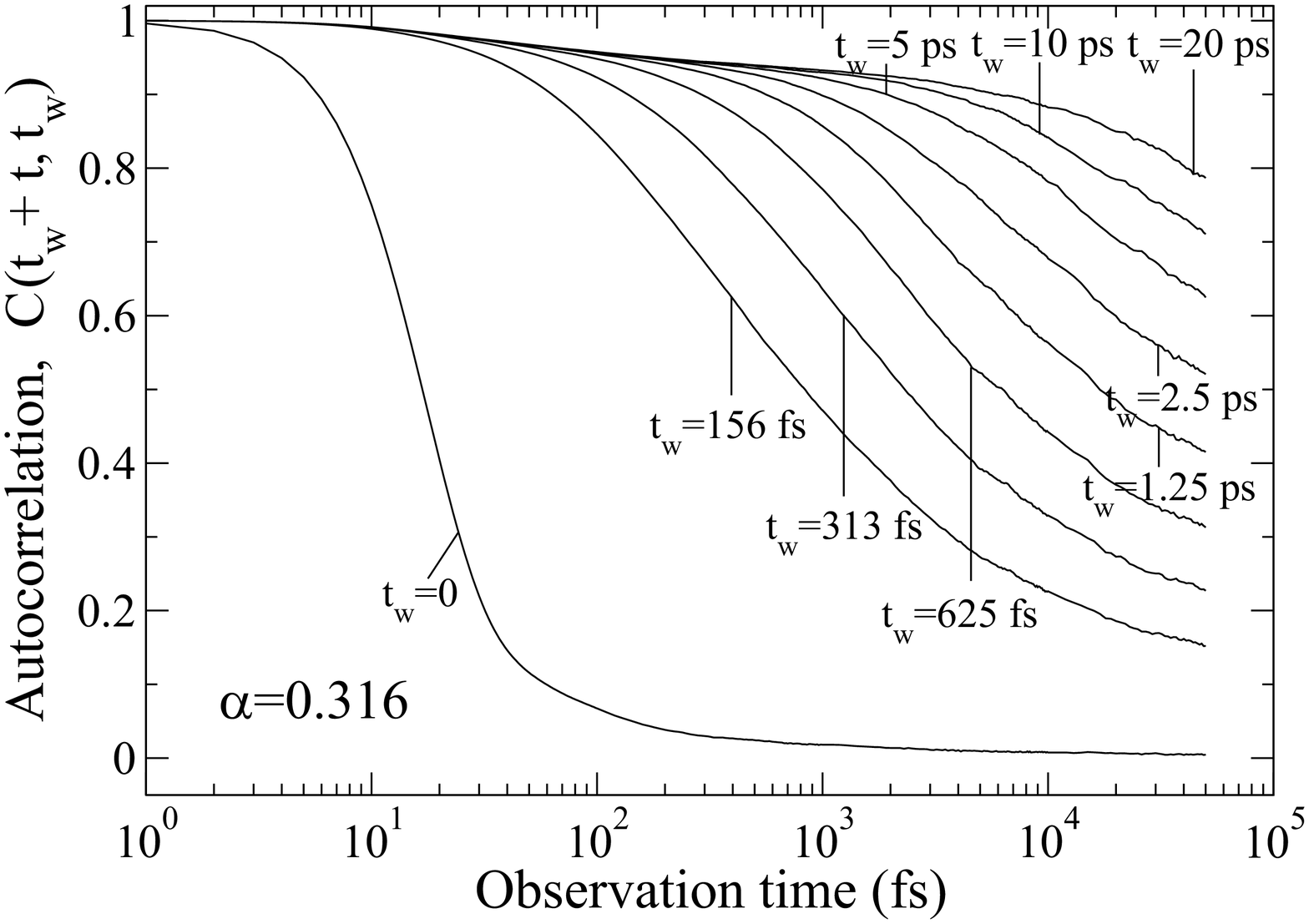}
\caption{Same as Fig.~\ref{fig:cumn1} but with damping $\alpha=0.1$.}
\label{fig:cumn3}
\end{figure}
\begin{figure}
\includegraphics*[width=0.45\textwidth]{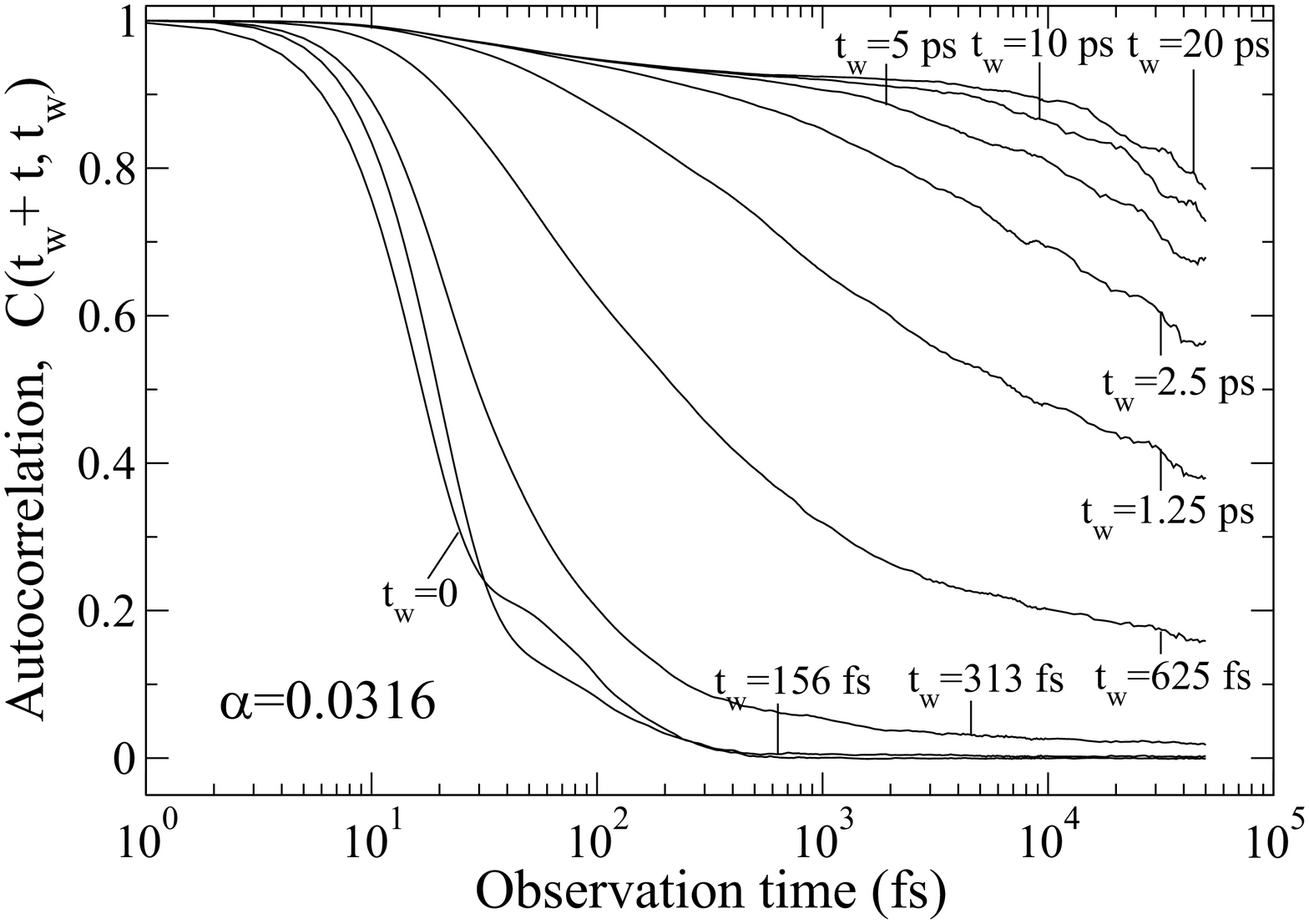}
\caption{Same as Fig.~\ref{fig:cumn1} but with damping $\alpha=0.316$.}
\label{fig:cumn4}
\end{figure}

\begin{figure}
\includegraphics*[width=0.45\textwidth]{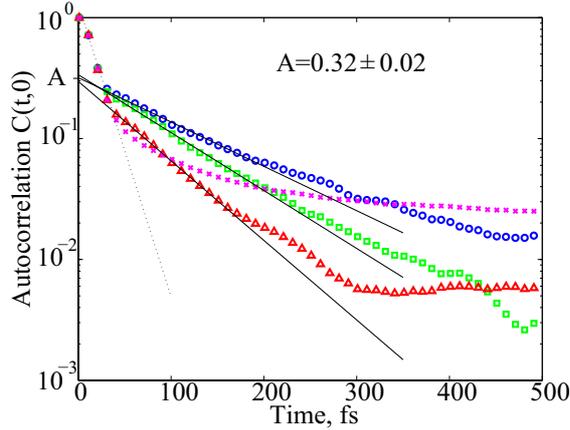}
\caption{Autocorrelation $C(t,t_w=0)$ for four values of the damping parameter:
$\alpha=0.01$ (circles), $\alpha=0.0316$ (boxes), $\alpha=0.1$ (triangles),
$\alpha=0.316$ (crosses). The dashed line is a linear fit to the points for
10~fs$\leq t\leq$ 40~fs; the slope is equal to $1/\tau_1$ of Eq.~\ref{eq:ct0}.
The slope of the solid lines corresponds to the damping relaxation rate
$1/\tau_2$ in Eq.~\ref{eq:ct0}.}
\label{fig:reltw0}
\end{figure}

In Figs.~\ref{fig:cumn1}-\ref{fig:cumn4} we show the spin glass
dynamics of CuMn for four different damping parameters,
$\alpha=0.01, 0.0316, 0.1, 0.316$ respectively. We plot the
autocorrelation function for logarithmically spaced waiting times.
Note that the time is given in femtoseconds, but the time step size
in the simulation is 0.01~fs. In all four cases, the autocorrelation
for $t_w$=0 illustrates the initial dynamics of the system right
after the quench. The behavior at short times ($t\lesssim 500$~fs)
is similar for all values of $\alpha$ and is shown in
Fig.~\ref{fig:reltw0} in log-linear scale. The evolution of the
autocorrelation function can be described here as a sum of two
exponents followed by a slower-than-exponential decay at larger
times. That is for $t\lesssim 200$~fs, we have
\begin{equation}
C(t,0)\approx (1-A)e^{-\frac{t}{\tau_1}}+A e^{-\frac{t}{\tau_2}},
\label{eq:ct0}
\end{equation}
where $A$ can be extracted from the the intersection point of
straight lines corresponding to the second exponent:
$A=0.32\pm0.02$. The bump in Figs.~\ref{fig:cumn1}-\ref{fig:cumn4}
on the curves for $t_w=0$ corresponds to the second term in
Eq.~\ref{eq:ct0}. The initial slope of the curves, $1/\tau_1$, is
independent of damping (see Fig.~\ref{fig:reltw0}) and temperature
(data not shown) and depends only on the details of the Hamiltonian
and initial spin distribution. When the initial distribution is
random, as is the case in present simulations, the drop of the
autocorrelation is dominated by the strong precessional motion of
the atomic spins in rapidly varying effective exchange fields. As
the directions of the effective fields initially are completely
randomly oriented, the angle between the atomic spin and its
effective field is on average large, resulting in a large
precessional torque on the atomic spins. The system gradually
relaxes by means of a damping torque on each atomic spin, with the
energy of the system dropping down from a high value of the random
spin configuration ("high-temperature" phase) to a value close to
the average energy for $T=10 K$.

The subsequent decay of the autocorrelation is associated with the
equilibration of spins in their local fields. Clearly, the rate of
this relaxation, $1/\tau_2$, depends strongly on the value of the
damping parameter and for this reason we refer to it as "damping
relaxation". As the rate $1/\tau_2$ diminishes with increasing
$\alpha$, the initial damping relaxation becomes more difficult to
identify. As it is seen in Fig.~\ref{fig:reltw0} for $\alpha=0.316$,
the crossover from the initial stage to non-exponential decay is
rather smooth and relaxation due to damping is indistinguishable.

The rate of the damping relaxation affects the behavior of the
autocorrelation function for waiting times much larger than the
value of $\tau_2$. From Fig.~\ref{fig:cumn1} one can see that for
$\alpha=0.01$, the curves fall on top of each other for waiting
times up to $t_w=625$ fs. This implies that by this moment, the
decay of the autocorrelation function has become time-translation
invariant. In fact, it seems that for this value of the damping
parameter, the system never enters the aging regime and the initial
relaxation phase crosses over directly to relaxation to the global
equilibrium for $t_w>5$ fs. On the other hand, at smaller values of
$\tau_2$, the aging behavior recovers (see
Figs.~\ref{fig:cumn3},\ref{fig:cumn4}) and the spin dynamics is
similar to the case of infinite damping. This implies that the
strength of damping is determined by the the ratio of the timescale
$\tau_2$ of the damping relaxation and the characteristic time of
detuning of local fields due to the motion of neighboring spins
contributing to these local fields.

\begin{figure}
\includegraphics*[width=0.45\textwidth]{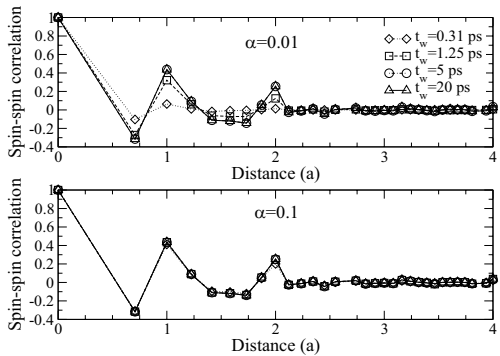}
\caption{The calculated spin-spin correlation function,\break
$\langle \mathbf{m}_i\cdot\mathbf{m}_j\rangle$, for $\alpha=0.01$
(upper panel) and $\alpha=0.1$ (lower panel).
The correlation function is plotted for four logarithmically spaced waiting times.
For $\alpha=0.1$, all four curves fall essentially on top of each other.}
\label{fig:overlap}
\end{figure}

To determine to what extent the system has equilibrated, one can
look at the evolution of the spin-spin correlation function
$g(\mathbf{r}_{ij})=\langle \mathbf{m}_i\cdot\mathbf{m}_j\rangle$.
In Fig.~\ref{fig:overlap} we plot $g(\mathbf{r}_{ij})$ as a function
of the distance $|\mathbf{r}_{ij}|$ between the spins for different
waiting times of the system. The correlation function is plotted
both for $\alpha=0.01$ (upper panel) and $\alpha=0.1$ (lower panel).
As expected from the autocorrelation, $g(\mathbf{r}_{ij})$ is seen
to evolve faster for the larger damping. It means that for
sufficiently large damping the system reaches the quasi-equilibrium
phase fast enough for the aging regime to establish.

\begin{figure}
\includegraphics*[width=0.45\textwidth]{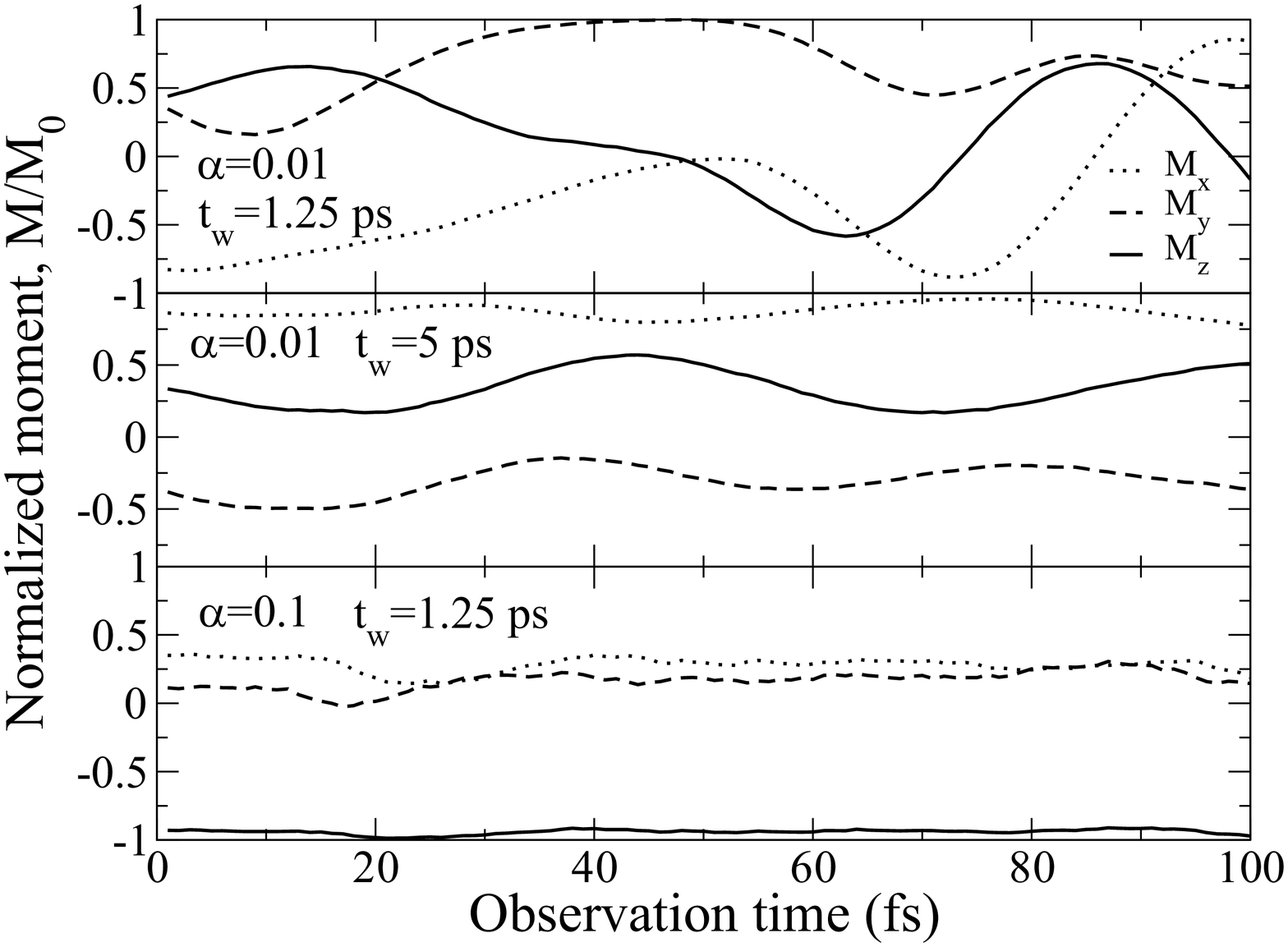}
\caption{The trajectory for one typical atomic spin at $t_w=1.25$~ps
for $\alpha=0.01$ (upper panel), at $t_w=5$~ps for $\alpha=0.01$ (middle panel)
and at $t_w=1.25$~ps for $\alpha=0.1$ (lower panel).}
\label{fig:traj}
\end{figure}

The non-equilibrium behavior seen at small waiting times for small
damping parameter values can be detected on the microscopic level by
observing the trajectories of randomly selected spins. In
Fig.~\ref{fig:traj} we plot the trajectory of a typical atomic spin
evolving during 100~fs (corresponding to a short time scale), for
$\alpha=0.01$ and $\alpha=0.1$, and for two different waiting times.
The upper panel shows a trajectory for $\alpha=0.01$ after a waiting
time of 1.25~ps. There is a large degree of precessional motion of
the atomic spin, confirming the conclusions which were drawn from
the autocorrelation that the system is still in the initial
relaxation phase at this waiting time. The middle panel shows the
same system after a waiting time of 5~ps, showing an atomic spin
with a much more stable spin direction. The spin is now either in
equilibrium or on the verge of entering equilibrium, although spin
motion is much more pronounced here than in the aging regime for the
system with $\alpha=0.1$. The lower panel shows the trajectory of an
atomic spin for $\alpha=0.1$ at a waiting time of 1.25~ps. The
system is here in the aging regime, as seen in Fig.~\ref{fig:cumn2},
and the atomic spin direction is stable on a time scale of 100~fs.

In the aging regime, the autocorrelation is characterized by an
initial reduction of the autocorrelation on to a plateau, similar to
what is seen for systems in equilibrium. A plateau is clearly seen
in Figs.~\ref{fig:cumn1}-\ref{fig:cumn4} for larger waiting times.
The position of the plateau depends on temperature and is related to
the spin-glass order parameter. More precisely,
\begin{equation}
\lim_{t \to \infty}\lim_{t_w \to \infty}C(t_w+t,t_w) = q_{\mathrm{EA}}
\end{equation}
in the macroscopic limit, and the Edward-Anderson order parameter,
$q_{\mathrm{EA}}$, is defined (again in the macroscopic limit) as
\begin{equation}
q_{\mathrm{EA}}=\frac{1}{N}\sum_{i}\left[\langle\mathbf{m}_i\rangle^{2}_{T}\right]_{\mathrm{av}},
\end{equation}
where $\langle\dots\rangle_{T}$ is thermal averaging. Following the
plateau, or the quasi-equilibrium phase, is the aging phase. The
crossover from one phase to another occurs at a time, $t_s$, when a
sudden drop of the autocorrelation takes place. The time $t_s$ can
be best identified as the maximum of the relaxation rate defined by
Eq.~\ref{eq:relax}.

\begin{figure}
\includegraphics*[width=0.45\textwidth]{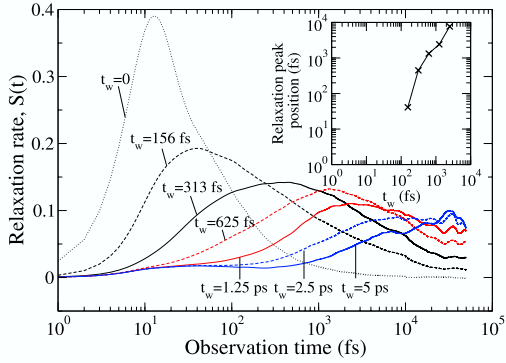}
\caption{Relaxation rate, $S(t)$, for the simulation in Fig.~\ref{fig:cumn3}.
The relaxation rate is plotted for the logarithmically spaced waiting times
$t_w=0$, $1.56\cdot 10^2$, $3.13\cdot 10^2$, $6.25\cdot 10^2$, $1.25\cdot 10^3$,
$2.5\cdot 10^3$, and $5\cdot 10^3$~fs. The inset shows the relationship between
the waiting time and the peak of the relaxation rate for the waiting times
$t_w=1.56\cdot 10^2$, $3.13\cdot 10^2$, $6.25\cdot 10^2$, $1.25\cdot 10^3$,
and $2.5\cdot 10^3$~fs. For the last waiting time the largest peak seems
to be a numerical artifact. Instead the second largest peak was chosen as input for the inset.}
\label{fig:relax}
\end{figure}

The relaxation rate for $\alpha=0.1$ and for a few waiting times is
plotted in Fig.~\ref{fig:relax}. The relaxation rate is calculated
by performing a derivative with respect to $\ln{t}$ of the
autocorrelation. Note that, the poorly defined peaks at the end of
the observation time ($t \sim 10^4$~fs) are artifacts of a smearing
scheme, used when calculating the derivative and which breaks down
close to the edge of the observation interval. In the inset we show
the position of the peak relaxation rate, or $t_s$, with respect to
the waiting time. As one can see, $t_s$ is slightly larger than
$t_w$, which is expected for the aging regime in a spin-glass
system. \cite{Zotev_2003} However, the total time window used in the
simulations does not allow to infer any definite form of the
dependence.

\subsection{Spin dynamics of the EA model for weak damping}

To investigate even further spin dynamics for small $\alpha$, we
have performed simulations of the EA model for $\alpha=0.01$ and for
different lattice sizes. Simulations have been performed on a cubic
lattice of different sizes $L\times L\times L$, where $L$=4, 8, 16,
and with random nearest neighbor exchange interactions drawn from a
Gaussian distribution with a standard deviation of 1 mRy. This is
typically the order of the exchange interaction. The freezing
temperature, $T_g$, is expected to be 25~K for this model (0.16
within the dimensionless model).\cite{Berthier2004}

\begin{figure}
\includegraphics*[width=0.45\textwidth]{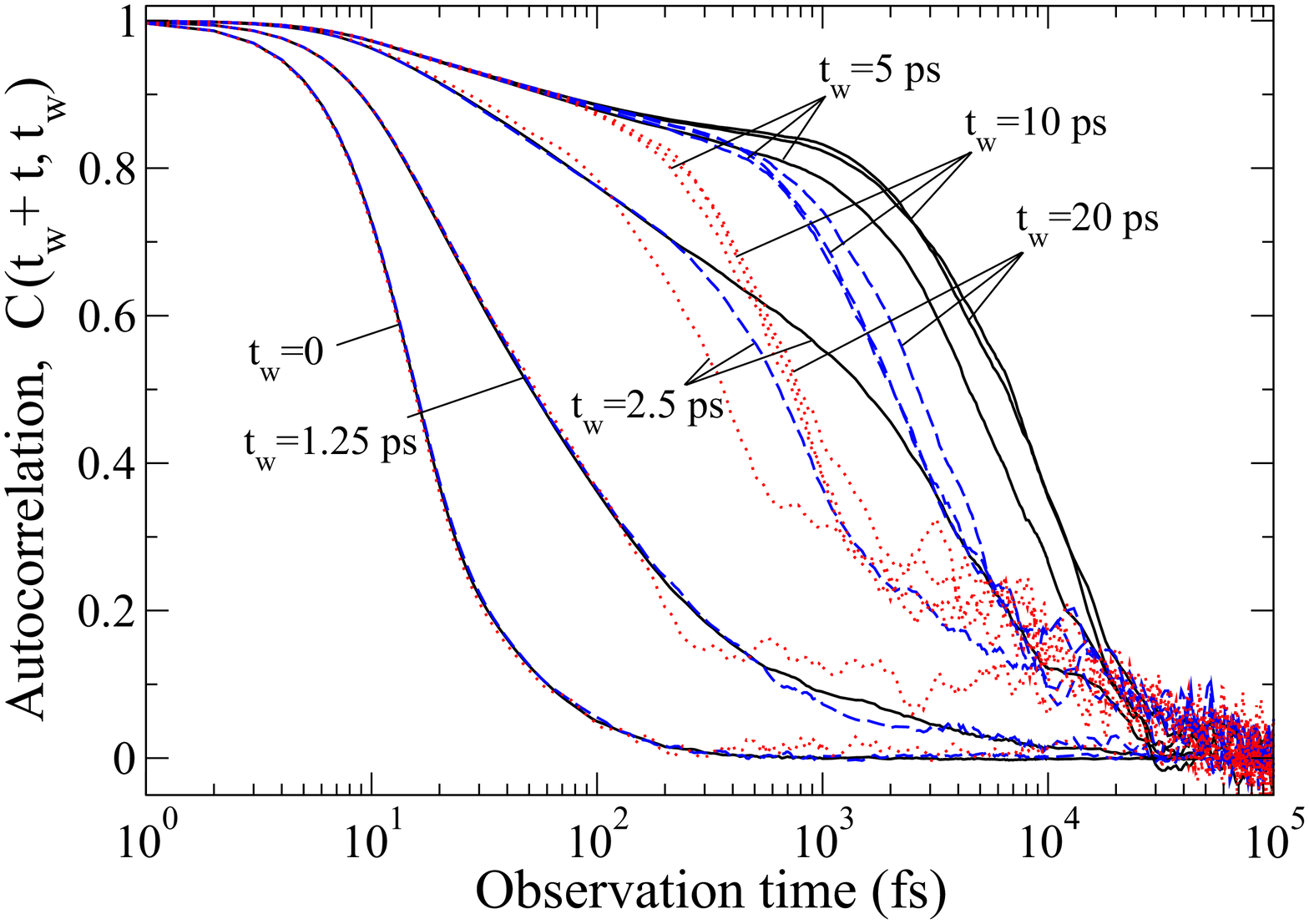}
\caption{Autocorrelation $C(t_w+t, t_w)$ calculated for the Edwards-Anderson model
after a quench from completely random spin orientations to T=10~K with $\alpha=0.01$.
Different colors signify different system sizes:
$4\times 4\times 4$ (red), $8 \times 8\times 8$ (blue), $16 \times 16\times 16$ (black).
For each system size the autocorrelation is presented (from left to right in the figure)
for the logarithmically spaced waiting times $t_w=0$,
$1.25\cdot 10^3$, $2.5\cdot 10^3$, $5\cdot 10^3$, $1\cdot 10^4$, and $2\cdot 10^4$~fs.}
\label{fig:ea}
\end{figure}

In Fig.~\ref{fig:ea}, we show the calculated autocorrelation for a
simulation of the Edwards-Anderson model. The simulated process is a
relaxation following a quench from completely random spin
orientations to 10~K (0.063 within the standard dimensionless
model). The autocorrelation with respect to observation time is
plotted for several logarithmically spaced waiting times. As in the
case of the CuMn simulations, averaging was performed over 10
different bond realizations, and for each bond configuration, 10
simulations with different initial random spin distributions and
different random number sequences in the Langevin equations have
been done.

There are three sets of curves in Fig.~\ref{fig:ea} for three
different system sizes. As it is seen, for the choice of the damping
parameter ($\alpha=0.01$) and system sizes ($L=4,8,16$), the aging
regime is very short or even non-present in these simulations. A
global equilibrium is reached very soon after the initial relaxation
has been accomplished. It is also worth noting that with the same
damping parameter, there is a strong similarity between the waiting
time dependence of the auto correlation function for the
16$\times$16$\times$16 EA model (Fig.~\ref{fig:ea} ) and the dilute
20$\times$20$\times$20 CuMn alloy (Fig.~\ref{fig:cumn1}). Moreover,
comparing the curves corresponding to the initial phase ($t_w=0$),
one can see that this initial phase is independent of the system
size.

Typically, a spin glass system enters the aging regime as soon as
local equilibrium conditions are being met. The dynamics proceeds by
a rearrangement of the magnetic order on a length scale
corresponding to a time scale of the order of the age of the system.
In this particular simulation, a pure aging regime can not be
identified as the system enters a global equilibrium soon after the
initial relaxation. In contrast to equilibrium, within the aging
regime, the autocorrelation should depend on the waiting time and
not on the system size. For the largest four waiting times in
Fig.~\ref{fig:ea}, we see the autocorrelation characterized by an
initial reduction on to a plateau followed by a large sudden
reduction to zero for different observation times depending on the
size of the system.

A global equilibrium is reached for finite systems, as the
correlation length reaches the size of the system. If there is aging
dynamics, it can no longer proceed. In equilibrium, thermal
fluctuations continue to govern the motion of the atomic spins.
Since there is no energy associated with a global rotation of the
system, the autocorrelation is reduced to zero in equilibrium. The
autocorrelation is now translationally invariant with respect to the
observation time, $t$. By comparing simulations on systems of
different size we see that by increasing the size of the system by a
factor 2, the equilibration time is increased approximately by a
factor $3$. This is due to the fact that the correlation length
grows logarithmically in time.

\section{Conclusions}
The investigation of spin dynamics based on the realistic spin-glass
model has been performed by solving the Langevin equations of
motion. The exchange parameters have been extracted from
first-principles DFT calculations, while the damping parameter has
been varied to study the influence of damping on the dynamics. The
simulations showed that below the spin-freezing temperature, $T_g$,
the system exhibits the aging behavior for large enough values of
the damping parameter, $\alpha$. In this case, the dynamics is very
similar to that obtained from corresponding Monte-Carlo simulations.

At weak damping, however, the behavior is different and can
basically be characterized by two regimes for small and large
waiting times, respectively. For waiting times, $t_w$, below some
certain value, the autocorrelation function does not depend on $t_w$
(i.e. it is time-translation invariant) and hence is the same as for
$t_w=0$. For waiting times above a certain limit, the
autocorrelation is also time-translation invariant but is
characterized by much slower decay. The time-translation invariance
of the autocorrelation suggests that a system of finite size reaches
equilibration faster at weaker damping. As a result, it becomes
clear that spin dynamics inside moderately sized domains in
spin-glasses can be strongly affected by damping.

\acknowledgments Financial support from the Swedish Foundation for
Strategic Research (SSF), Swedish Research Council (VR), the Royal
Swedish Academy of Sciences (KVA), Liljewalchs resestipendium and
Wallenbergstiftelsen is acknowledged. Calculations have been
performed at the Swedish national computer centers UPPMAX, HPC2N and
NSC.


\end{document}